\documentclass[aip,apl,twocolumn,preprintnumbers,amsmath,amssymb]{revtex4}

\usepackage{graphicx}
\usepackage[colorlinks=true,citecolor=blue]{hyperref}

\def\ket#1{\mathinner{|{#1}\rangle}}

\begin{document}
    %\linenumbers
    \title{Phonon Mediated Off-Resonant Quantum Dot-Cavity Coupling}
    \author{Arka Majumdar}
    \email{arkam@stanford.edu}
    \author{Yiyang Gong}
    \author{Erik D. Kim}
    \author{Jelena Vu\v{c}kovi\'{c}}
    \affiliation{$^1$E.L.Ginzton Laboratory, Stanford University, Stanford, CA, $94305$\\}

    %\date{\today}
\begin{abstract}
A theoretical model for the phonon-mediated off-resonant coupling
between a quantum dot and a cavity, under resonant excitation of
the quantum dot, is presented. We show that the coupling is caused
by electron-phonon interaction in the quantum dot and is enhanced
by the cavity. We analyze recently observed resonant quantum dot
spectroscopic data by our theoretical model.
\end{abstract}
\maketitle
%\section{Introduction}
One of the interesting recent developments in cavity quantum
electrodynamics (CQED) experiments with quantum dots (QDs) coupled
to semiconductor micro-cavities is the observation of off-resonant
dot-cavity coupling. This unusual phenomenon is observed both in
photoluminescence studies, where an above-band laser generates
carriers that incoherently relax to recombine at the QD frequency
\cite{article:hen07, article:press07, article:finley08} and also
under resonant excitation of the QD or the cavity
\cite{article:majumdar09,article:michler09}. The coupling observed
via photoluminescence is attributed to several phenomena including
pure QD dephasing \cite{article:auffves09}, the electron-phonon
interaction \cite{article:arakawa09,
article:electron_phonon_cqed}, multi-exciton complexes
\cite{article:immamoglu09} and charges in the vicinity of the QD
\cite{article:prb09}. To isolate the role of phonons in
off-resonant QD-cavity coupling, studies employing resonant
excitation of the QD are preferable as they avoid possible
complications arising from multi-excitonic complexes and nearby
charges generated in above band pumping. Resonant excitation of a
QD coupled to an off-resonant cavity has been recently used to
perform QD spectroscopy, enabling the observation of power
broadening of the QD line-width and saturation of the cavity
emission
\cite{article:majumdar10,article:michler10,article:erik10}.
However, there is presently no theoretical model accounting for
off-resonant dot-cavity coupling for the case of resonant QD
excitation. Although off-resonant coupling can be modeled by
introducing a phenomenological incoherent cavity pumping mechanism
\cite{incoherent_pumping}, such a treatment masks the actual
physical phenomenon responsible for the coupling. Without an
explicit coherent driving term in the system Hamiltonian, the
resonant QD spectroscopy results cannot be explained.

In this Letter, we theoretically model the phonon-mediated
interaction between a cavity mode and an off-resonant QD under
resonant excitation of the QD. We first model the coupling via
pure QD dephasing. Then, we propose a new model where the
phonon-mediated coupling is enhanced by the presence of the
cavity. We compare these two models and find that they provide
qualitatively similar signatures in terms of experimental resonant
QD spectroscopic studies, such as power broadening and saturation
of a resonantly driven QD \cite{article:majumdar10,
article:michler10}. However, in the newly proposed model, the
coupling is maintained at very large QD-cavity detunings ($\sim 3$
meV), as observed in the recent experiments
\cite{article:majumdar10}. We also observe the signature of the
inherent asymmetry between phonon emission and absorption rate in
the simulated QD spectroscopy results, depending on whether the QD
is red or blue detuned from the cavity.

%\section{Theory with Pure QD Dephasing}
The dynamics of a coherently driven QD (with a ground state
$\ket{g}$ and an excited state $\ket{e}$) coupled to an
off-resonant cavity mode is governed by the Hamiltonian $H$ (in a
frame rotating at the driving laser frequency)
\begin{equation}
H=\Delta\omega_c a^\dag a +\Delta\omega_a \sigma^\dag
\sigma+ig(a^\dag \sigma -a\sigma^\dag)+\Omega(\sigma+\sigma^\dag)
\end{equation}
where, $a$ and $\sigma$ are the annihilation and lowering
operators for the cavity mode and the QD, respectively;
$\Delta\omega_c = \omega_c-\omega_l$ and $\Delta\omega_a =
\omega_a-\omega_l$ are the cavity and dot detunings from the
driving laser, respectively; $\Delta=\omega_a-\omega_c$ is the
QD-cavity detuning; $\Omega$ is the Rabi frequency of the driving
laser and is proportional to the laser field amplitude and $g$ is
the coherent interaction strength between the QD and the cavity.

In this coupled system, there are two independent mechanisms for
energy dissipation: cavity decay and QD dipole decay. The system
losses can be modeled by the Liouvillian, and the Master equation
describing the dynamics of the lossy system is given by
\begin{equation}
\label{Maseq} \frac{d\rho}{dt}=-i[H,\rho]+ 2\kappa
\mathcal{L}[a]+2\gamma \mathcal{L}[\sigma]
\end{equation}
where $\rho$ is the density matrix of the coupled QD-cavity
system, $2\gamma$ and $2\kappa$ are the QD spontaneous emission
rate and the cavity population decay rate, respectively. We
neglect any non-radiative decay of the QD exciton.
$\mathcal{L}[D]$ is the Lindblad operator corresponding to a
collapse operator $D$ and is given by:
\begin{equation}
\mathcal{L}[D]= D\rho D^\dag-\frac{1}{2}D^\dag D
\rho-\frac{1}{2}\rho D^\dag D
\end{equation}
In addition, phonons in the solid state system destroy the
coherence of the exciton. This is generally modeled by adding an
additional incoherent decay term
$2\gamma_d\mathcal{L}[\sigma^\dag\sigma]$ to the Master equation,
where $2\gamma_d$ is the pure dephasing rate of the QD. This term
destroys the polarization of the QD without affecting the
population of the QD. The dissipation of the QD polarization and
population $(\sigma_z=[\sigma^\dag,\sigma])$ is given by the
following mean-field equations:
\begin{eqnarray}
  \frac{d\langle\sigma\rangle}{dt} &=& -(\gamma+\gamma_d) \langle\sigma\rangle \\
  \frac{d\langle\sigma_z\rangle}{dt}&=&-2\gamma(1+\langle\sigma_z\rangle)
\end{eqnarray}
Hence, the linewidth of the QD, at the zero excitation power
limit, is given by $2(\gamma+\gamma_d)$. However, in this model,
the effect of phonons is embedded in the phenomenological pure
dephasing rate $\gamma_d$, which affects only the QD, and does not
include any cavity effects.

%\section{Theory with cavity assisted phononic coupling}
We now propose a different model for off-resonant dot-cavity
coupling, where the phonon-mediated coupling strength is affected
by both the cavity and the QD.
\begin{figure}
\centering
\includegraphics[width=3in]{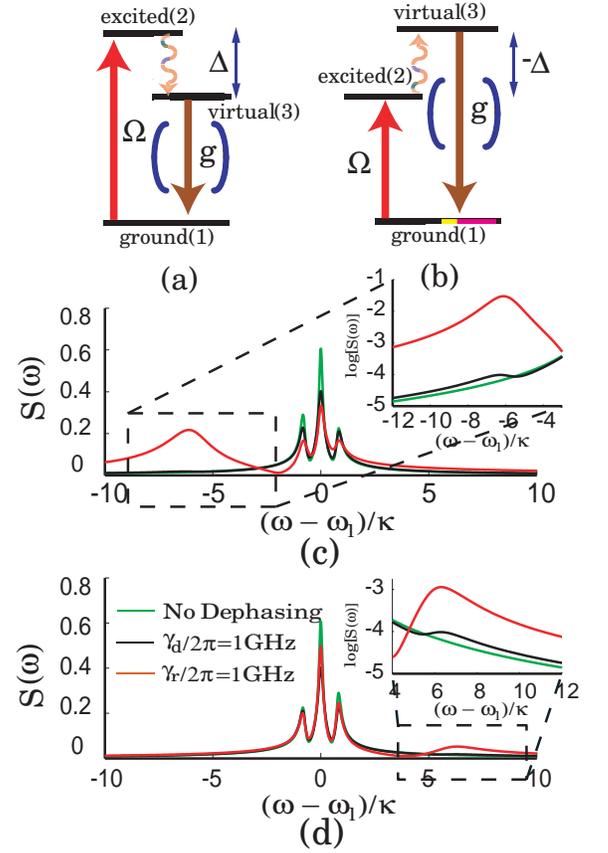}
\caption{(a),(b) Level diagram of the coupled QD/cavity system for
blue (a) and red (b) detuned QD relative to the cavity resonance.
A laser drives the quantum dot (transition between ground state
$(1)$ and excited state $(2)$) resonantly. The excited state $(2)$
can decay via two paths: the first is by direct decay back to the
ground state $(1)$ via the spontaneous emission of the QD; the
second is by indirect decay via the emission (Fig.
\ref{Figure_1_level_theory}a) or absorption (Fig.
\ref{Figure_1_level_theory}b) of a phonon (transition $(2)$ to
$(3)$) and subsequent emission of a photon at the cavity frequency
(transition $(3)$ to $(1)$). (c), (d) Emission $S(\omega)$ as a
function of frequency $\omega$ for a QD resonantly driven by a
laser ($\Delta \omega_a=0$). The plot (c) corresponds to a QD that
is blue detuned, and (d) corresponds to a QD that is red detuned
from the cavity. Three different cases are considered: first,
without any dephasing; second with pure dephasing and third with a
cavity enhanced phonon process (our newly introduced model). We
observe the Mollow triplet at the QD frequency in all three cases.
The inset of (c) and (d) show an enlarged view of the emission at
the cavity frequency. We observe no off-resonant cavity emission
without pure dephasing, or our newly introduced cavity enhanced
phonon process. For the simulation we assume
$g/2\pi=\kappa/2\pi=20$ GHz; $\gamma/2\pi=1$ GHz; $\Omega/2\pi=6$
GHz; QD-cavity detuning $\Delta/2\pi=\pm6$ GHz.}
\label{Figure_1_level_theory}
\end{figure}
The effect of phonons can be modeled by replacing the pure
dephasing term with two additional incoherent decay terms in the
Master equation. For blue-detuned QD (Fig.
\ref{Figure_1_level_theory} a) we have:
%\begin{equation}
%\label{Maseq} \frac{d\rho}{dt}=-i[H,\rho]+ 2\kappa
%\mathcal{L}[a]+2\gamma \mathcal{L}[\sigma]\\
%+2\gamma_r\bar{n}\mathcal{L}(\sigma^\dag
%a)+2\gamma_r(\bar{n}+1)\mathcal{L}(\sigma a^\dag)
%\end{equation}
\begin{equation}
\label{phonon_eq}
\begin{split}
  \frac{d\rho}{dt} =& -i[H,\rho]+ 2\kappa \mathcal{L}[a]+2\gamma
\mathcal{L}[\sigma] \\
   & +2\gamma_r\bar{n}\mathcal{L}(\sigma^\dag
a)+2\gamma_r(\bar{n}+1)\mathcal{L}(\sigma a^\dag)
\end{split}
\end{equation}
where, $2\gamma_r$ is an effective decay rate of the QD exciton
states via the emission of a phonon and a photon at the
off-resonant cavity frequency. $\bar{n}$ is the average number of
phonons at the dot-cavity detuning frequency $(\Delta)$ present in
the system at thermal equilibrium with the reservoir at a
temperature $T$, and is given by
\begin{equation}
\bar{n}(\Delta,T)=\frac{1}{e^{\hbar\Delta/k_BT}-1}
\end{equation}
The analysis for a QD red detuned from the cavity (Fig.
\ref{Figure_1_level_theory} b) can be carried out in a similar
manner by replacing the final two terms of Eqn.\ref{phonon_eq}
with $2\gamma_r\bar{n}\mathcal{L}(\sigma a^\dag)$ and
$2\gamma_r(\bar{n}+1)\mathcal{L}(\sigma^\dag a)$.

The decay term $\sigma^\dag a$ denotes the annihilation of a
cavity photon and excitation of the QD, while the term $\sigma
a^\dagger$ denotes the creation of a cavity photon and collapse of
the QD to its ground state, accompanied by the creation (or
annihilation) of phonons to compensate for the QD-cavity frequency
difference. Only the second process is important for observing
cavity emission under resonant excitation of the dot. We also note
that the observation of QD emission under resonant excitation of
the cavity \cite{article:majumdar09} can be modeled in the same
way by changing the coherent driving term from
$\Omega(\sigma+\sigma^\dag)$ to $\Omega(a+a^\dag)$. In this
situation, the collapse operator $\sigma^\dag a$ is important.
Similar decay channels have been proposed to model cavity assisted
atomic decay \cite{article:ray_pra}. A detailed derivation of
these incoherent terms can be found in the Supplementary
Materials. We call this process a cavity-enhanced phonon process.

We first consider the case where the QD is blue detuned from the
cavity (Fig. \ref{Figure_1_level_theory} a). The qualitative
nature of the dissipation of the QD polarization and population
can be determined by the mean-field equations (assuming
$\bar{n}=0$, i.e., at the zero temperature limit):
\begin{eqnarray}
\label{eq_lw_phonon}
  \frac{d\langle\sigma\rangle}{dt} &=& -\gamma \langle\sigma\rangle-\gamma_r(1+\langle a^\dag a\rangle)\langle\sigma\rangle \\
  \frac{d\langle\sigma_z\rangle}{dt}&=&-2\gamma(1+\langle\sigma_z\rangle)-2\gamma_r(1+\langle a^\dag a\rangle)(1+\langle\sigma_z\rangle)
\end{eqnarray}
We notice that, unlike the pure dephasing case, both the QD
population and polarization are affected by the cavity enhanced
phonon process. The linewidth of the QD at the zero excitation
power limit is given by $2\gamma+2\gamma_r(1+\langle a^\dag
a\rangle)$, which is qualitatively different from what we expect
from a pure QD dephasing model (as the presence of cavity photons
affects the QD linewidth). We emphasize that for both models, the
QD linewidths obtained from the mean-field equations neglect the
modification resulting from cavity QED effects in the presence of
the cavity and hence the predicted linewidths are only recovered
at large QD-cavity detunings.

%\section{Calculation of Spectra}
The resonance fluorescence of the system is given by the power
spectral density (PSD) of the coupled system. As we collect the
fluorescence primarily from the cavity, the PSD is calculated as
the Fourier transform of the cavity field auto-correlation:
\begin{equation}
S(\omega)=\int_{-\infty}^{\infty}\langle a^\dag(\tau)a(0)\rangle
e^{-i\omega\tau}d\tau
\end{equation}
To determine the two time correlation functions and, subsequently,
the PSD, we use the quantum regression theorem
\cite{book:quan_noise}.

We simulate the coupled system using numerical integration
routines provided in the quantum optics toolbox \cite{qotoolbox}
with realistic system parameters $\kappa/2\pi = g/2\pi =20$ GHz
and $\gamma/2\pi = 1$ GHz. The off-resonant coupling is observed
both for a strongly $(g>\kappa)$ and a weakly coupled QD to the
cavity $(g<\kappa)$. Figure \ref{Figure_1_level_theory} c shows
the numerically calculated resonance fluorescence spectra obtained
from the cavity under resonant excitation of a blue detuned QD for
three different cases: no dephasing, pure QD dephasing and
phonon-induced exciton decay at a bath temperature of $4$ K. We
first note that no emission is observed at the cavity frequency in
the absence of pure dephasing or cavity enhanced phonon process
(see inset of Fig. \ref{Figure_1_level_theory} c). Though the pure
dephasing ($\gamma_d/2\pi=1$ GHz) and phonon-induced decay
($\gamma_r/2\pi=1$ GHz) cases both show off-resonant cavity
emission, the latter shows enhanced cavity emission. In all three
cases, we also observe the Mollow triplet at the QD frequency, as
expected from QD resonance fluorescence \cite{book:scully_book}.
The Mollow side-band closer to the cavity is enhanced compared to
the other side-band causing an asymmetric triplet. Fig.
\ref{Figure_1_level_theory} d shows the similar spectra for a
red-detuned QD. Our model predicts that the emission at cavity
frequency is considerably smaller for a red detuned QD compared to
a blue-detuned one as expected at finite temperature (as the
process for a red-detuned dot relies on phonon absorption, while
for a blue-detuned dot relies on phonon emission).

\begin{figure}
\centering
\includegraphics[width=3.5in]{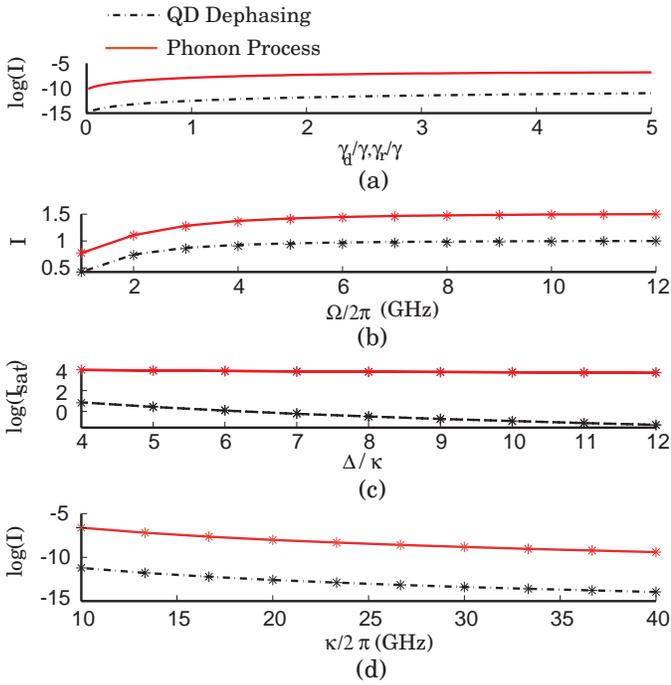}
\caption{Resonance fluorescence $I$ collected from the cavity for
a blue-detuned QD ($\Delta>0$). (a) $log (I)$ as a function of the
rates $\gamma_d$ and $\gamma_r$ (for two models, respectively).
$\Delta/\kappa=10$.(b) Normalized cavity fluorescence as a
function of the Rabi frequency $\Omega$ of the laser for two
models. Saturation of the cavity emission is observed. The
fluorescence values are not in scale. (c) Dependence of the
saturated cavity emission $I_{sat}$ on the dot-cavity detuning
$\Delta$. (d) $log(I)$ as a function of the cavity linewidth
$\kappa$. (Parameters used for all the simulations are
$g/2\pi=\kappa/2\pi=20$ GHz; $\gamma/2\pi=1$ GHz. For Figs.
\ref{Fig2_saturation_emission} a, c and d, $\Omega/2\pi=4$ GHz and
for Figs. \ref{Fig2_saturation_emission} b,c and d,
$\gamma_d/2\pi=\gamma_r/2\pi=1$ GHz.)
}\label{Fig2_saturation_emission}
\end{figure}

%\section{Saturation of Emission and Dependence on the Detuning}
We then perform simulations to observe how the emission collected
at the cavity frequency depends on the dephasing and
phonon-induced decay rates, as well as on the field of the driving
laser resonant with the QD $(\Delta\omega_a=0)$. We observe that
the cavity emission $I$ increases almost linearly with the pure QD
dephasing rate $\gamma_d$ [Fig. \ref{Fig2_saturation_emission} a],
but exhibits a nonlinear dependence on the rate $\gamma_r$ when
the coupling is enhanced by the presence of the cavity. Fig.
\ref{Fig2_saturation_emission} b shows the cavity fluorescence as
a function of the laser Rabi frequency $\Omega$. For both models,
the cavity fluorescence $I$ follows a saturation curve
\begin{equation}
I=I_{sat}\frac{\tilde{P}}{1+\tilde{P}}
\end{equation}
where, $\tilde{P}\propto\Omega^2$ and $I_{sat}$ is the saturated
cavity emission intensity. As noted previously in the article, the
cavity emission is higher when the process is enhanced by the
presence of the cavity. Furthermore, we investigate the dependence
of the saturation emission intensity $I_{sat}$ as a function of
the QD-cavity detuning $\Delta$ (Fig.
\ref{Fig2_saturation_emission} c). $I_{sat}$ falls of as
$1/\Delta^2$ with the detuning $\Delta$ when the dot-cavity
coupling is modeled as a pure dephasing process. However, when the
coupling is modeled as a cavity enhanced phonon process, the
saturation intensity exhibits a diminished dependence on detuning
$\Delta$ (estimated to be $\sim 1/\Delta^{0.25}$). Hence, one may
observe off-resonant coupling for larger detunings when the phonon
process is enhanced by the cavity. Fig.
\ref{Fig2_saturation_emission} d shows $log(I)$ as a function of
the cavity decay rate $\kappa$, for a fixed detuning of
$\Delta/2\pi=200$ GHz. For both models, the emission falls off as
$1/\kappa^2$, signifying that the off-resonant coupling does not
depend on the overlap between the QD and the cavity spectra.

\begin{figure}
\centering
\includegraphics[width=3.5in]{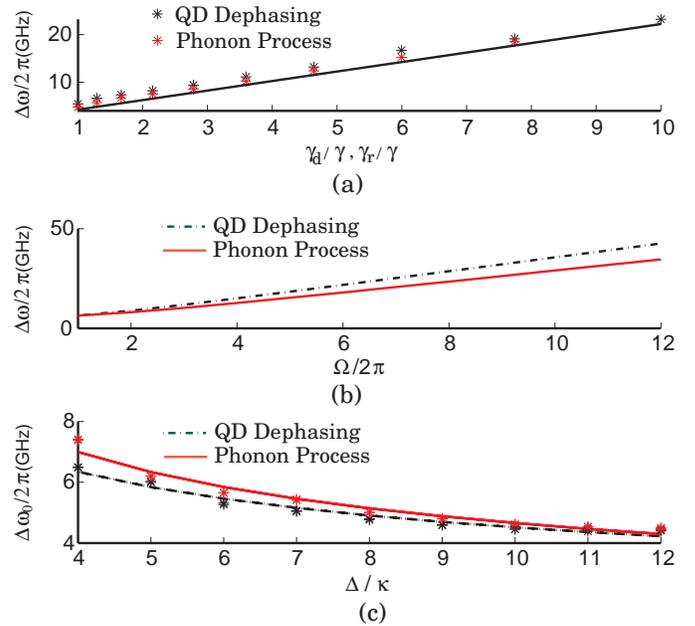}
\caption{Linewidth of the blue detuned QD relative to the cavity
$(\Delta>0)$ as measured by monitoring the off-resonant cavity
emission (similar to the experiments in Ref.
\cite{article:majumdar10}). Results obtained from two models are
presented. (a) QD linewidth as a function of the rates $\gamma_d$
and $\gamma_r$ for pure dephasing and the cavity enhanced phonon
process, respectively. In both cases, $\Omega/2\pi=1$ GHz and
$\Delta=12\kappa$. The solid black line shows the theoretical
estimates of the QD linewidth when the laser excitation power is
very low and the QD is not significantly perturbed by the cavity.
(b) Power broadened linewidth vs laser Rabi frequency $\Omega$ for
QD-cavity detuning $\Delta/2\pi=6$ GHz (for the case of pure QD
dephasing and the cavity enhanced phonon process). (c) Dependence
of the intrinsic QD linewidth $\Delta\omega_0$ on dot-cavity
detuning $\Delta$. $\Delta\omega_0/2\pi$ approaches
$2(\gamma+\gamma_d)/2\pi$ or $2(\gamma+\gamma_r)/2\pi$ ((both
chosen to be $4$ GHz) with large $\Delta$. (Parameters used for
all the simulations are: $\kappa/2\pi=g/2\pi=20$ GHz;
$\gamma/2\pi=1$ GHz.)} \label{Fig3_linewidth_broadening_detuning}
\end{figure}

%\section{Power Broadening and Linewidth with detuning}
We now measure the QD line-width $\Delta\omega$ monitoring the
cavity emission, while scanning the laser wavelength across the QD
resonance, similar to the experiments in Ref.
\cite{article:majumdar10}. We observe that at very low excitation
power $\Omega/2\pi=1$ GHz and large QD-cavity detuning
$\Delta/2\pi=12\kappa$, the linewidths of the QD are very close to
the theoretical linewidth in the absence of the cavity (shown by
the solid black line in Fig.
\ref{Fig3_linewidth_broadening_detuning} a). At a constant
QD-cavity detuning and laser excitation power, the QD line-width
increases with increasing $\gamma_d$ and $\gamma_r$ (Fig.
\ref{Fig3_linewidth_broadening_detuning} a). We observe broadening
of the line-width with increasing laser power (Figs.
\ref{Fig3_linewidth_broadening_detuning} b). The power broadened
QD linewidth $\Delta\omega$ is fit with the model
$\Delta\omega=\Delta\omega_0\sqrt{1+\tilde{P}}$, where
$\Delta\omega_0$ is the intrinsic line-width of the QD and
$\tilde{P}$ is obtained from the fit to the saturation of the
cavity emission \cite{article:majumdar10}. The theoretical model
does not reproduce the additional power-independent broadening of
the QD \cite{article:majumdar10} and we believe that this extra
broadening may result from QD spectral diffusion
\cite{fluctuating_environment_PRB}. We analyze the intrinsic QD
linewidth $\Delta\omega_0$ (without power broadening, i.e.,
obtained from plots in Figs.
\ref{Fig3_linewidth_broadening_detuning} b at $\Omega=0$ limit) as
a function of the dot-cavity detuning $\Delta$ for two different
models (Figure \ref{Fig3_linewidth_broadening_detuning} c). We
observe that at large detuning, $\Delta\omega_0$ approaches the
unperturbed QD linewidth $2(\gamma+\gamma_d)$ and
$2(\gamma+\gamma_r)$, respectively. We fit empirical models of
$\Delta^{-\alpha}$ to the intrinsic linewidths for the two models
and find that with pure QD dephasing, $\Delta\omega_0$ falls off
more slowly ($\alpha \simeq 0.4$) compared to the cavity enhanced
coupling ($\alpha \simeq 0.7$). The weak dependence of the
intrinsic QD linewidths on the dot-cavity detuning shows that the
off-resonant cavity does not perturb the QD significantly.

\begin{figure}
\centering
\includegraphics[width=3.5in]{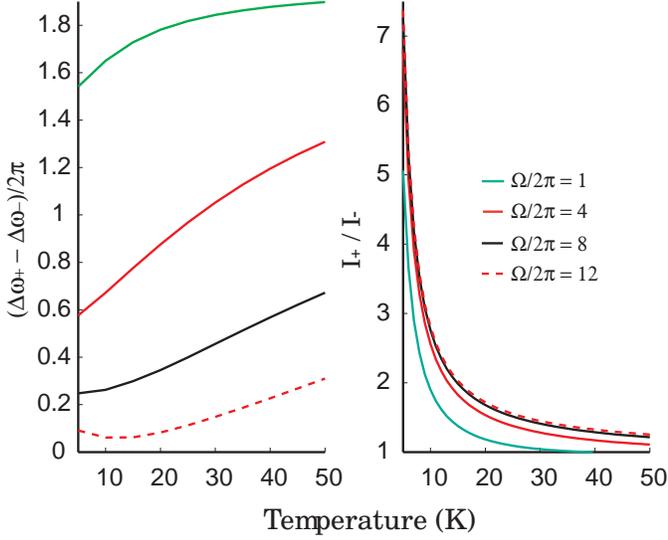}
\caption{(a) The difference in QD linewidths measured via
collected emission through off-resonant cavity, for a blue
($\Delta \omega_+$) and a red ($\Delta \omega_-$) detuned QD for
different values of the excitation laser Rabi frequency $\Omega$
as a function of the bath temperature $T$. (b) Ratio of the cavity
intensity for an off-resonant QD blue $(I_+)$ and red $(I_-)$
detuned from the cavity, as a function of the bath temperature
$T$. For all the simulations, the absolute value of QD-cavity
detuning is kept at $10\kappa$. } \label{Figure_4_asymmetry}
\end{figure}

However, the results are dramatically different for  a red-detuned
QD. This is a result of the fact that the rate of the incoherent
process involving the operator $\sigma a^\dag$ is different
depending on whether the QD is red or blue detuned from the cavity
(because at any temperature, the rates of absorption and emission
of phonons are different). This asymmetry is manifested in both
the emission collected at the cavity resonance and in the QD
spectroscopy result. Fig. \ref{Figure_4_asymmetry} a shows the
difference in the QD linewidths as a function of the bath
temperature $T$ for different driving laser Rabi frequencies
$\Omega$, when the QD is red or blue detuned from the cavity by
same absolute value. We observe that the difference in linewidth
is higher when the QD is weakly driven and hence is not power
broadened. Full quantum optical simulations show that the
linewidth depends on the bath temperature, but reaches the maximum
value of $2\gamma_r$ at a higher temperature. Fig.
\ref{Figure_4_asymmetry} b shows the ratio of the cavity emission
as a function of the bath temperature for different driving laser
Rabi frequencies $\Omega$. The difference in cavity intensity is
maximum at lower bath temperature and is almost zero at higher
temperature.

%\section{Conclusion}
In conclusion, we have proposed a theoretical model for the
off-resonant dot-cavity coupling under resonant excitation of the
QD. Introduction of an incoherent decay channel (referred to as
the cavity enhanced phonon process, which is different from pure
dephasing) shows that the phonon-mediated dot-cavity coupling is
enhanced by the presence of the cavity. By comparing the
power-broadening and saturation of the QD between the cavity
enhanced phonon process and pure QD dephasing, we found that the
dot-cavity coupling is enhanced in the former case and is observed
for a larger QD-cavity detuning. Our model can also be used to
explain asymmetry in the spectroscopy results for a QD blue and
red detuned from the cavity. We believe that such an off-resonant
dot-cavity coupling can be used as an efficient read-out cannel
for resonant QD spectroscopy and for QD-spin manipulation.

The authors acknowledge financial support provided by the National
Science Foundation and Army Research Office. A.M. was supported by
the Stanford Graduate Fellowship (Texas Instruments fellowship).

\section{Supplementary Materials}
\subsection{Derivation of Decay Terms}
We use the level diagram as shown in Fig.
\ref{Figure_1_level_theory} a to model the effect of phonons
explicitly. The Hamiltonian of the system is given by
\begin{equation}
H=H_0+H_I
\end{equation}
where,
\begin{equation}
H_0=\omega_1|1\rangle\langle 1|+\omega_2|2\rangle\langle
2|+\omega_3|3\rangle\langle 3|+\omega a^\dag a+\sum_j\nu_jb_j^\dag
b_j
\end{equation}
and
\begin{equation}
H_I=g_v(a|3\rangle\langle 1|+a^{\dag}|1\rangle\langle 3|)+\sum_j
g^j_{23}(b_j^\dag|3\rangle\langle 2|+b_j|2\rangle\langle 3|)
\end{equation}
where, $|i\rangle\langle i|$ is the population operator for
$i^{th}$ level; $a$ is the annihilation operator for the cavity
mode; $b_j$ is the annihilation operator for a phonon in the
$j^{th}$ mode. $\omega_i$ , $\omega$ and $\nu_j$ are the
frequencies of the $i^{th}$ energy level, cavity resonance and a
phonon in the $j^{th}$ mode. $g_v$ signifies the interaction
strength between the cavity and the virtual transition and
$g_{23}^j$ is the interaction strength between the QD exciton and
an $j^{th}$ mode phonon. We note that this interaction Hamiltonian
is valid only for the level structure as in Fig.
\ref{Figure_1_level_theory} a, where the cavity is at lower energy
than the QD. For the situation in Fig. \ref{Figure_1_level_theory}
b (where the cavity is of higher energy compared to the QD), the
interaction Hamiltonian $H_I$ is given by
\begin{equation}
H_I=g_v(a|3\rangle\langle 1|+a^{\dag}|1\rangle\langle 3|)+\sum_j
g^j_{23}(b_j|3\rangle\langle 2|+b_j^\dag|2\rangle\langle 3|)
\end{equation}
In the following derivation, we will use the situation shown in
Fig. \ref{Figure_1_level_theory} a.

If we define the QD resonance frequency as $\omega_a$, then
$\omega_a=\omega_2-\omega_1$; and the cavity frequency is given by
$\omega_c=\omega_3-\omega_1$. Then the QD-cavity detuning is given
by $\Delta=\omega_2-\omega_3$. Defining
$\sigma_{ij}=|i\rangle\langle j|$, we can write
%\begin{equation}
%\dot{\sigma_{13}}=-i[\sigma_{13},H_0+H_I]=-i\omega_c\sigma_{13}-iga(\sigma_{11}-\sigma_{33})-i\sum_jg^j_{23}b_j^\dag\sigma_{12}
%\end{equation}
\begin{eqnarray*}
  \dot{\sigma_{13}} &=& -i[\sigma_{13},H_0+H_I] \\
   &=& -i\omega_c\sigma_{13}-ig_va(\sigma_{11}-\sigma_{33})-i\sum_jg^j_{23}b_j^\dag\sigma_{12}
\end{eqnarray*}
Similarly,
%\begin{equation}
%\dot{\sigma_{23}}=-i[\sigma_{23},H_0+H_I]=i\Delta\sigma_{23}-iga\sigma_{21}-i\sum_jg^j_{23}b_j^\dag(\sigma_{22}-\sigma_{33})
%\end{equation}
\begin{eqnarray*}
  \dot{\sigma_{23}} &=& -i[\sigma_{23},H_0+H_I] \\
   &=& i\Delta\sigma_{23}-ig_va\sigma_{21}-i\sum_jg^j_{23}b_j^\dag(\sigma_{22}-\sigma_{33})
\end{eqnarray*}
Separating the slow and the fast components of the operators, we
can write
\begin{eqnarray}
  \sigma_{13} &=& \tilde{\sigma}_{13}e^{-i\omega_c t}\\
  \sigma_{23} &=& \tilde{\sigma}_{23}e^{-i\omega_j t}\\
  \sigma_{12} &=& \tilde{\sigma}_{12}e^{-i(\omega_c-\omega_j) t}\\
   a &=& \tilde{a}e^{-i\omega_c t}\\
   b_j^\dag &=& \tilde{b}_j^\dag e^{-i\omega_j t}
\end{eqnarray}
Hence the equations governing the dynamics of the system can be
written as
\begin{equation}
\dot{\tilde{\sigma}}_{13}=-ig_v\tilde{a}(\tilde{\sigma}_{11}-\tilde{\sigma}_{33})-i\sum_jg^j_{23}\tilde{b}_j^\dag\tilde{\sigma}_{12}
\end{equation}
and
\begin{equation}
\dot{\tilde{\sigma}}_{23}=i(\Delta-\omega_j)\tilde{\sigma}_{23}-ig_v\tilde{a}\tilde{\sigma}_{21}-i\sum_jg^j_{23}\tilde{b}_j^\dag(\tilde{\sigma}_{22}-\tilde{\sigma}_{33})
\end{equation}
As level $3$ is a virtual level, it is never populated. Hence by
adiabatic elimination, using
$\dot{\tilde{\sigma}}_{13}=\dot{\tilde{\sigma}}_{23}=0$, we obtain
\begin{equation}
\label{eq_ad1}
\tilde{\sigma}_{23}=\frac{g_v\tilde{a}\tilde{\sigma}_{21}+\sum_jg^j_{23}\tilde{b}_j^\dag(\tilde{\sigma}_{22}-\tilde{\sigma}_{33})}{\Delta-\omega_j}
\end{equation}
and
\begin{equation}
\label{eq_ad2}
\tilde{\sigma}_{12}=\frac{-g_v\tilde{a}(\tilde{\sigma}_{11}-\tilde{\sigma}_{33})}{\sum_jg^j_{23}\tilde{b}_j^\dag}
\end{equation}
Using these values, we can find the interaction Hamiltonian. The
first term $g_v(a\sigma_{31}+a^\dag\sigma_{13})$ denotes the
coherent dynamics. The second term, which signifies the effect of
phonons, can be written as (using the Eqns. \ref{eq_ad1} and
\ref{eq_ad2})
\begin{eqnarray}
  H_{ph} &=& \sum_j g^j_{23}(b_j^\dag\sigma_{32}+b_j\sigma_{23})  \\
  & =& \sum_j g^j_{23}(\tilde{b}_j^\dag\tilde{\sigma}_{32}+\tilde{b}_j\tilde{\sigma}_{23})  \\
   &=& \sum_j\frac{g^j_{23}g_v}{\Delta-\omega_j}(\tilde{b}_j^\dag
   \tilde{a}^\dag\tilde{\sigma}_{12}+\tilde{b}_j \tilde{a}
   \tilde{\sigma}_{21})\\
   & &+\sum_j\frac{(g^j_{23})^2}{\Delta-\omega_j}(\tilde{\sigma}_{22}-\tilde{\sigma}_{33})(\tilde{b}_j^\dag \tilde{b}_j^\dag+\tilde{b}_j\tilde{b}_j)
\end{eqnarray}
The second term involves two-phonon processes which are less
likely. If we neglect them, we can model the effect of phonons as
follows:
\begin{equation}
H_{ph}= \sum_j\frac{g^j_{23}g_v}{\Delta-\omega_j}(\tilde{b}_j^\dag
\tilde{a}^\dag\tilde{\sigma}_{12}+\tilde{b}_j \tilde{a}
\tilde{\sigma}_{21})
\end{equation}
We can write this Hamiltonian as
\begin{equation}
H_{ph}= (\tilde{a}^\dag
\tilde{\sigma}_{12}\tilde{\Gamma}^\dag+\tilde{a}\tilde{\sigma}_{21}\tilde{\Gamma})
\end{equation}
where, the operator $\tilde{\Gamma}$ can be written as
\begin{equation}
\tilde{\Gamma}=\sum_j\frac{g_{23}^jg_v}{\Delta-\omega_j}\tilde{b}_j
\end{equation}
and
\begin{equation}
\Gamma=\sum_j\frac{g_{23}^jg_v}{\Delta-\omega_j}b_je^{-i\omega_jt}
\end{equation}
To obtain the familiar Lindblad term, we take the partial trace of
the correlation between the reservoir operators over the reservoir
variables. The correlation is given by
\begin{equation}
\langle\Gamma^\dag(t')\Gamma(t)\rangle_R=\sum_j\left|\frac{g_{23}^jg_v}{\Delta-\omega_j}\right|^2e^{-i\omega_j(t-t')}\langle
b_j^\dag b_j\rangle_R
\end{equation}
As the operators $b_j$ are bosonic and the system is in thermal
equilibrium with a bath at temperature $T$, using the relation
\begin{equation}
\langle b_j^\dag
b_j\rangle_R=\bar{n}(\omega_j,T)=\frac{1}{e^{\frac{\hbar\omega_j}{k_BT}}-1}
\end{equation}
we find
\begin{equation}
\langle\Gamma^\dag(t')\Gamma(t)\rangle_R=\sum_j\left|\frac{g_{23}^jg_v}{\Delta-\omega_j}\right|^2e^{-i\omega_j(t-t')}\bar{n}(\omega_j,T)
\end{equation}
and
\begin{equation}
\langle\Gamma(t')\Gamma^\dag(t)\rangle_R=\sum_j\left|\frac{g_{23}^jg_v}{\Delta-\omega_j}\right|^2e^{-i\omega_j(t-t')}(\bar{n}(\omega_j,T)+1)
\end{equation}
From the correlation, we find that the phonons with frequency
$\Delta$ (corresponding to the difference between levels
$|2\rangle$ and $|3\rangle$, i.e., QD-cavity mode detuning), have
the maximum contribution in the interaction Hamiltonian. In the
Born-Markov approximation, we can model the electron-phonon
interaction (for Fig. \ref{Figure_1_level_theory} a) as an
incoherent decay process by adding two extra terms to the Master
equation: $2\gamma_r\bar{n}\mathcal{L}(\sigma^\dag a)$ and
$2\gamma_r(\bar{n}+1)\mathcal{L}(\sigma a^\dag)$, $\gamma_r$ being
the effective decay rate of the excited QD state and is given by
\begin{equation}
\gamma_r=\frac{1}{2}\sum_j\left|\frac{g_{23}^jg_v}{\Delta-\omega_j}\right|^2
\end{equation}
For the situation shown in Fig. \ref{Figure_1_level_theory} b, the
decay terms are given by $2\gamma_r\bar{n}\mathcal{L}(\sigma
a^\dag)$ and $2\gamma_r(\bar{n}+1)\mathcal{L}(\sigma^\dag a)$. We
note that the different rates in both cases are due to an inherent
asymmetry between the absorption and emission rates of the
phonons.
\subsection{Derivation of the Mean Field Equations}
To find the mean field equations for an operator $A$ from the
Master equation, we used the following relation:
\begin{equation}
\frac{d\langle A
\rangle}{dt}=\frac{d}{dt}Tr[A\rho]=Tr\left[A\frac{d\rho}{dt}\right]
\end{equation}

For the cavity enhanced phonon process, the mean field equations
for a non-zero $\bar(n)$ is given by (when the QD is blue detuned
from the cavity)
\begin{eqnarray}
  \frac{d\langle\sigma\rangle}{dt} &=& -\gamma \langle\sigma\rangle-\gamma_r(1+\langle a^\dag a\rangle)\langle\sigma\rangle \\
  & &\gamma_r\bar{n}(1+2\langle a^\dag a\rangle)\langle\sigma\rangle\\
  \frac{d\langle\sigma_z\rangle}{dt}&=&-2\gamma(1+\langle\sigma_z\rangle)\\
  & & -2\gamma_r(1+\bar{n})(1+\langle a^\dag a\rangle)(1+\langle\sigma_z\rangle)
\end{eqnarray}

When the QD is red detuned from the cavity, the mean field
equations are:
\begin{eqnarray}
  \frac{d\langle\sigma\rangle}{dt} &=& -\gamma \langle\sigma\rangle-\gamma_r\langle a^\dag a\rangle\langle\sigma\rangle \\
  & &\gamma_r\bar{n}(1+2\langle a^\dag a\rangle)\langle\sigma\rangle\\
  \frac{d\langle\sigma_z\rangle}{dt}&=&-2\gamma(1+\langle\sigma_z\rangle)\\
  & & -2\gamma_r \bar{n}(1+\langle a^\dag a\rangle)(1+\langle\sigma_z\rangle)
\end{eqnarray}

We note that while deriving these mean-field equations, we assume
that the cavity and QD operators are uncorrelated and write
\begin{equation}
\langle a^\dag a \sigma\rangle=\langle a^\dag a \rangle\langle
\sigma\rangle
\end{equation}
Under weak driving, the approximation holds very well. However, in
full quantum optical simulations, we do not make any assumptions.
\bibliography{NRDC_bibl}
\end{document}